\documentclass[12pt]{article}
\usepackage[]{epsfig}
\usepackage{graphicx}
\usepackage{amsfonts}
\usepackage{color}
\usepackage{color}

\newcommand{\be}{\begin{equation}}
\newcommand{\ee}{\end{equation}}
\newcommand{\bea}{\begin{eqnarray}}
\newcommand{\eea}{\end{eqnarray}}

\begin{document}

\title{Building Gauge Theories: The Natural Way}
\author{C.A. Garc\'{\i}a Canal, F.A. Schaposnik\thanks{Associated to CICBA}\\
Instituto de F\'{\i}sica La Plata, CCT La Plata, CONICET and \\
  Departamento de F\'{\i}sica,
Facultad de Ciencias Exactas, \\ Universidad Nacional de La Plata\\
CC 67, 1900 La Plata, Argentina} 
\date{}
\maketitle

\begin{abstract}

The construction of a gauge field theory for elementary particles usually starts
by promoting global invariance of the matter action to a local one, this in turn
implying the introduction of gauge fields. We present here a procedure that runs the other way:
starting from the action for gauge fields, matter is gauge invariantly coupled on the basis of Lorentz
invariance and charge conservation. This natural approach   prevents
using the concept of particles separated from gauge fields that mediate interactions.

\end{abstract}

Gauge invariance was introduced by H.~Weyl \cite{weil} in 1919 in an attempt of unifying electrodynamics
and general relativity. In his original proposal, gauging meant a space dependent change in the
space-time unit of length together with the rescaling of the ``compensating'' or connection fields,
the gauge potentials.  Weyl's original proposal   was  incorrect but, nevertheless,
the concept of gauge invariance survived as a
central symmetry of Maxwell equations, extremely
useful in the development of electrodynamics, particularly at the quantum level.
At present, electromagnetic interactions of matter mediated by
photons can be understood arising from the requirement of local Abelian
gauge invariance in the theory. This idea was generalized by Yang
and Mills \cite{YM} to the case of non-Abelian local symmetries in a work that can   be
considered as the starting point of the far-reaching analogy between
electromagnetism and the other fundamental forces.
The theoretical trail that was opened in 1954 has had a tremendous
success as   nowadays
 the fundamental interactions are
described by local gauge theories  \cite{MS}.

Weyl's idea was presented long before its own time becoming today a
fundamental ingredient of our present understanding of the fundamental interactions of Nature.
It is important to stress that all these advances  are based upon the basic concepts of gauge fields
(the electromagnetic potentials and their non-Abelian generalizations) and the corresponding group of gauge transformations.
The natural mathematical framework for this approach  is
  given by the theory of fiber bundles  \cite{DM}.

{Once one recognizes the connection between the electromagnetic field strength
  and the gauge field   covariant
 derivative, it becomes manifest that gauge invariance  is the paramount property of electrodynamics.
 Already in the non-relativistic quantum
description of a charged particle in an electromagnetic field, gauge
invariance imposes the introduction  a phase factor in the particle
wave function in order to maintain gauge covariance at the
Hamiltonian level. As it is well known the phase factor is directly
associated to the covariant derivative and in this way it becomes
clear how   gauge covariance is transmitted from the field strength
to   the particle wave function.}

{The standard way for constructing a gauge theory for matter fields seems to go in the
opposite direction. Indeed, one starts by
first  choosing on physical grounds an
appropriate unitary Lie group $G$ as the matter symmetry group   and proposes an
action governing the  dynamics. The action
should have a global invariance under $G$, i.e., the action should
be invariant under {constant} phase transformations associated to the elements
$g \in G$ and this implies the conservation of a Noether charge.}

The second step in the usual construction of gauge invariant actions is to promote the global invariance to
a local one, as in
Yang-Mills proposal,
so that $g \to g(x) \in \bar G$.  Here we write $\bar G$ to indicate
that now the group corresponds to local gauge transformations  {(i.e. the associated phases are space-time dependent)}.
In
order to have a gauge invariant {matter field} action, it is then mandatory to
introduce  gauge fields that
compensate, via their gauge transformation, the changes  produced   by
the local phase changes. As a result, the kinetic part of the matter Lagrangian turns to be written in terms of
covariant derivatives. Now,  once gauge fields are introduced, it becomes clear that
a kinetic term  for them should be included. Finally,
 under rather general  assumptions for the behavior of $g(x)$ at infinity, the conserved Noether charge remains
unchanged with respect to the global case.

~

It is the aim of this note  to show that it is possible and natural  to invert the
process of theory building described above. Namely, the idea is
 to avoid as starting point the ``decree" that promotes  the  originally  global  phase transformations of the matter Lagrangian to local ones. In our view, the proposal we present is
closer to  Weyl's ideas  and their modern geometrical
interpretation being tightly connected with the physical idea of interactions.
Our  main idea is to show that it is possible   to build the theory of fundamental
interactions starting from the gauge fields, the interaction
mediators. Then, when sources are added,  one obtains
the corresponding matter-gauge field theory by only imposing  the  local gauge
invariance and the Lorentz invariance of Maxwell equations, or their non-Abelian generalizations.
To illustrate these ideas we begin with electrodynamics and
afterwards we present a general non-Abelian case.

~

Let us start from the Maxwell equations. In the absence of sources,
  the ``physical'' pair of
Maxwell equations  read
\be
\partial_\mu F^{\mu\nu} =0
\label{M0}
\ee
with the  field strength $F_{\mu\nu}(x)$   defined in terms of the $U(1)$
gauge field $A_\mu(x)$ as
 \be F_{\mu\nu}(x) = \partial_\mu A_\nu(x) -
\partial_\nu A_\mu(x)
\ee
{(The other pair is of course given by the Bianchi identity)}.

{Local} gauge transformations,   related to the fact that $\vec{A}$ has a
definite curl (the magnetic field $\vec{B}$), but its divergence is
not fixed are the natural symmetry of Maxwell equations.
Such transformations read
 \be A_\mu(x) \to
A_\mu^\Lambda(x) = A_\mu(x) +
\partial_\mu \Lambda(x) \ee
and leave the field strength $F_{\mu\nu}$  unchanged so that equation (\ref{M0})
 is  gauge invariant.
 Maxwell equations can be derived from the Lagrangian density
\be
L_M = -\frac14 F_{\mu\nu}(x)F^{\mu\nu}(x)
\ee
 which is Lorentz invariant and gauge invariant.

In the presence of an external source $j^\mu_{ext}$, eq.(\ref{M0})
should be modified to
\be
\partial_\mu F^{\mu\nu} = e j^\nu_{ext}
\label{sou}
\ee
with $e \in \mathbb{R}$, the coupling associated to
the source. {Within a Lagrangian formulation,} the natural and simplest
Lorentz invariant term coupling   the source to the gauge field which leads
to eq.(\ref{sou}) is
\be L_{int} = eA_\mu(x) j^\mu(x)_{ext}
\label{li} \ee Moreover, if gauge invariance of the  Lagrangian \be
L = L_M + L_{int} \ee is to be maintained,
the external current should not change under gauge transformations
 \be
j_{ext}^\mu(x) \to {j^{ \mu\;\Lambda}_{ext}}(x)
= j^\mu_{ext}(x) \label{c1}\ee
and satisfy
 \be
 \partial_\mu j_{ext}^\mu(x)= 0 \label{c2} \ee
Condition (\ref{c1}) is the natural one for an  external
(non-dynamical for the moment) current.  Being the field strength
antisymmetric,  eq.(\ref{c2}) is forces by Maxwell equations
 (\ref{sou}).  All this
 ensures that the Lagrangian (\ref{li}) changes at most
as a total derivative under gauge transformations.

{The Maxwell equations are a consequence of
relativistic invariance, or better said, their structure is largely
prescribed by Lorentz symmetry. In the same token one can anticipate
the way matter should be coupled to gauge fields. Just to visualize
our main idea for theory building let} us now consider a dynamical
Dirac fermion $\psi(x)$ at the origin of the four-vector current
coupled to the gauge field.  We call it $j^\mu(x)$ to distinguish it
from the the previous external current and consider the (most
economic) case of a bilinear combination of fermions. Lorentz
invariance and gauge invariance
  guide the construction. Primo, Lorentz invariance forces  to
write the vector as
\be j^\mu(x) = \bar \psi(x) \gamma^\mu \psi(x)
\ee
where $\bar \psi = \psi^\dagger \gamma^0$. Secundo, gauge
invariance of $j^\mu$ (a requirement analogous to (\ref{c1}))  imposes
\be
\psi(x) \to \psi^\Lambda(x) = \exp(iq\Lambda(x)) \psi
\label{11}
 \ee with $q
\in \mathbb{R}$.

In order to make such   fermions a dynamical field one should include a
kinetic energy term. The simplest one is provided by the Dirac free
Lagrangian
\be L_D = i \bar \psi(x) \gamma^\mu\partial_\mu \psi(x)
\ee The total Lagrangian
\be L = L_M + L_D + L_{int}
 \ee
 will be
gauge invariant provided one identifies $q$ in (\ref{11}) with $e$ in (\ref{sou}). Then,  by means of the covariant derivative, $D_\mu$,
\be
D_\mu =  \partial_\mu - eA_\mu
\ee
 Lagrangian $L$  can be written in
the usual form
 \be
L= - \frac14 F_{\mu\nu}F^{\mu\nu} + \bar\psi\, \gamma^\mu D_\mu  \,\psi
 \ee
 yielding the so-called ``minimal electromagnetic
coupling". ~

The extension to the non Abelian case is straightforward.  We consider  gauge fields $A_\mu(x) =
A_\mu^a(x) t^a$ taking values in the Lie algebra   of  $SU(N)$ with
generators $t^a$.
Under a gauge transformation the gauge field changes as \be A_\mu(x)
\to A_\mu^\Lambda(x) = g^{-1}(x) A_\mu(x) g(x) - \frac{i}e g^{-1}(x)
\partial_\mu g(x) \ee with \be g(x) = \exp\left(i \Lambda(x)\right)
\ee
and the gauge phase is defined as $\Lambda(x) = \Lambda^a(x)t^a$.
Infinitesimally one has
\be
A_\mu(x) \to A_\mu^\Lambda(x)  =  A_\mu + \frac1e D_\mu [A]\Lambda(x)
+ {\cal O}(\Lambda^2)
\ee
where $D_\mu[A]$ is the covariant derivative now defined as
\be
D_\mu[A] = \partial_\mu + ie[A_\mu(x),~\,]
\ee
  transforming according to

\be
D_\mu[A^\Lambda] = g^{-1}(x) D_\mu[A] g(x)
\ee
The non-Abelian field strength $F_{\mu\nu} = F_{\mu\nu}^a t^a$  is defined
as
 \be F_{\mu\nu} = \partial_\mu A_\nu -
\partial_\nu A_\mu + e[A_\mu.A_\nu]
\ee
and the
Maxwell equations (\ref{M0}) are naturally extended to Yang-Mills equations
\be
D_\mu[A] F^{\mu\nu} = 0
\label{YMD}
\ee
Under gauge transformations the field strength changes covariantly,
\be
F_{\mu\nu}(x) \to F_{\mu\nu}^\Lambda(x) = g^{-1}(x) F_{\mu\nu} g(x)
\ee
and so does eq.(\ref{YMD}). Of course, this
equation  can be derived from the
 Yang-Mills Lagrangian
\be
L_{YM} = - \frac12 {\rm tr} F_{\mu\nu}F^{\mu\nu}
\ee

~

As for the Maxwell case, let us now introduce an external
source $j^\mu_{ext}$ in the Yang-Mills equations.  One then has
\be
D_\mu F^{\mu\nu} = e j^\nu_{ext}
\label{J} \ee
 where, for consistency,    the current should take
values in the Lie algebra of $SU(N)$
\be
 j^\mu(x)_{ext}= j_{ext}^{\mu a}(x) t^a
\ee

In order to derive eq.(\ref{J}) from a Lagrangian, the natural Lorentz
invariant interaction term
to add is
\be
L_{int} = e \,{\rm tr} \,{A_\mu j^\mu_{ext}}
\ee
Gauge invariance of the Lagrangian
\be
L = L_{YM} + L_{int}
\ee
will be guaranteed provided the external source changes according to
\be
 j_{ext}^\mu(x) \to j_{ext}^{\mu \Lambda}(x) = g^{-1}(x) j^\mu_{ext}(x) g(x) \label{J1}
 \ee
 and satisfies
 \be
  D_\mu[A] j_{ext}^\mu(x) = 0
  \label{J2}
\ee
Again, in order to make  the source $j^\mu_{ext}$ in
eq. (\ref{J}) dynamical, one should introduce Dirac fermions. Following the same
lines as in the Abelian case, one  defines a   current
$j^\mu$ taking values in the Lie algebra of $SU(N)$
 \be j^{\mu a}(x) = \bar \psi^i  \gamma^\mu
T^a_{ij} \psi^j \; , \;\;\; i,j=1,2,\ldots,N \label{cvv} \ee where
for simplicity we have taken fermions in the fundamental
representation of $SU(N)$ with generators $T^a$. The transformation law \be \psi(x) \to
\psi^\Lambda (x)   = \exp(iq \Lambda(x)) \psi(x) \ee ensures that
current (\ref{cvv}) changes as in (\ref{J1}) provided $q=e$.

Adding the natural kinetic energy term for fermions one ends up with a
total Lagrangian of the form
\be
L = - \frac14   F_{\,\mu\nu}^a F^{a \mu\nu} +
 \bar\psi_i (i\!\,{\not\!\partial} \delta_{ij}- e\!\,{\not\!\! A^aT^a_{ij}})\psi_j
\ee defining a gauge field theory of $SU(N)$.

One should note that in this non-Abelian case, the matter current
$j_\mu$ is, according to eq.(\ref{J2}), covariantly conserved and
then it does not lead, by itself, to a conserved charge. It is only
when the gauge Noether current
 $({j^{gauge}})_\mu^a = f^{abc}F_{\mu\nu}^b A^{\nu c}$ associated to the Yang-Mills Lagrangian
is included that
  a conserved charge  $J_\mu = (j^{gauge})_\mu + j_\mu$ can be defined, verifying $\partial^\mu
  J_\mu = 0$.

~

As stated in  the introduction, the usual way in which one builds up the Lagrangian of
 charged matter fields coupled to gauge fields
is to promote the global unitary symmetry of the matter
Lagrangian to a local (gauge) symmetry, this requiring the
introduction of a gauge field. We have followed here the  opposite
way: we started from the pure gauge field Lagrangian and considered
the constraints imposed by Lorentz and gauge invariance when
coupling a matter source ending with the same Lagrangian and
conserved Noether current. Moreover, our presentation is parallel
to the geometric approach to gauge theories where the starting point
is to define a connection in a principal fiber bundle fiber   and then
introduce matter fields as sections in the associated  vector bundle \cite{FB}.

As we have shown, this procedure accepts generalizations to any gauge group.
It does not take the matter field Lagrangian as a starting point but matter is gauge invariantly coupled
on the basis of Lorentz invariance and charge conservation.

~

Partially supported by ANPCyT, Argentina. We warmly thank
illuminating comments by F. Cornet, F. del Aguila and J. S\'anchez
Guill\'en

\end{document}